\documentstyle[prd,preprint,aps]{revtex}
\begin{document}
\draft
\preprint{
\begin{tabular}{r}
REVISED
\\
JHU-TIPAC 940023
\\
DFF-215/12/94
\end{tabular}
}
\title{A New Scale-Dependent Cosmology
with the Generalized Robertson--Walker Metric and Einstein
Equation}
\author{C. W. Kim\footnote{E-mail cwkim@jhuvms.hcf.jhu.edu},
A. Sinibaldi\footnote{E-mail sinibaldi@fi.infn.it}
and J. Song\footnote{E-mail jhsong@rowland.pha.jhu.edu}}
\address{* \ddag Department of Physics and Astronomy\\
The Johns Hopkins University\\
Baltimore, MD 21218, U.S.A.\\
and}
\address{* \dag Dipartimento di Fisica, Univ. di Firenze\\
I.N.F.N. Sezione di Firenze, Firenze, Italy}
\maketitle

\begin{abstract}
\setlength{\baselineskip}{.5cm}
Based on the observed increase of $\Omega _0$
as a function of cosmic scale, the Robertson--Walker metric and
the Einstein equation are generalized so that
$ \Omega_0$, $H_0$, and
the age of the Universe, $t_0$, all become functions of cosmic
scales at which we
observe them.  The calculated local (within our galaxy) age of the
Universe is about 18 Gyr, consistent with the ages of the oldest
stars and globular clusters in our galaxy, while the ages at
distant
scales are shorter than the local age, solving the age puzzle.
It is also shown that $H_0$ increases as scale
increases, qualitatively consistent  with the recent observations.
\end{abstract}


In this Letter, we propose a new scale-dependent cosmology in which
the Hubble constant($H_0$), $\Omega_{0}(\equiv \rho_{0}/\rho_{c})$, and
the age of the Universe($t_0$), all become scale-dependent.
The need  of such a model was recently discussed by one of us (CWK [1])
in an attempt to resolve the recent two cosmological conflicts
by incorporating the running (scale-dependent) gravitational constant
as suggested by the Asymptotically-Free Higher-Derivative
(AFHD) quantum gravity [2] into cosmology.
The first conflict refers to the disagreement between the two
recent measurements [3,4] of  $H_{0}$,
the second being the discrepancy between
the calculated age of the Universe of about 8 Gyr,
based on the larger of the two observed $H_0$,
and the measured ages  of 14 to 18 Gyr  for the
oldest stars and globular clusters in our galaxy [5].

In [1], the treatment was entirely relied on the result of the
AFHD quantum gravity.
Here, we present a completely different approach which is purely
classical without resort to quantum gravity and is motivated
by the observed increase of $\Omega_{0}$ as a function of cosmic scale [6].
Our model is based on the following two ansatzs.
\begin{itemize}
\item
The recent epoch of the matter-dominated Universe is described by
the metric
\begin{equation}
d\tau^2 = dt^2-R^2(t,r)[dr^2+r^2d\Omega^2]~~ ,
\end{equation}
where $R(t,r)$ is the generalized scale factor which depends on
both  $t$ and $r$. This metric manifestly violates the Cosmological Principle.
\item
The equation of gravity is given by a generalized version of the
Einstein equation, i.e.,
\begin{equation}
R^{\mu \nu}-\frac{1}{2}g^{\mu \nu}R=-8\pi G T^{\mu \nu},
\end{equation}
with a constraint $[G T^{\mu \nu}]_{; \nu} =0$,
where a  semicolon denotes a covariant derivative.
An immediate consequence is that a product of the new gravitational $constant$
$G$ and the energy density $\rho$ is a function of $r$ as well as $t$.
$G$ becomes the Newton's constant, $G_N$, when $r=0$ at present epoch.
\end{itemize}

The first ansatz is motivated by the observation
that the present Universe with large scale structures and with changing
$\Omega_{0}$ as a function of scale cannot be isotropic and
homogeneous as implied by the Cosmological Principle.
Rather, the present Universe is locally
inhomogeneous but still appears approximately isotropic to us.
Since two observers at different locations
observe $different$ isotropic $\rho(r)$ (after some angular
average and ignoring local variations), the Universe cannot be
reduced to be homogeneous.
(Recall that for a Universe with perfect fluid, isotropy implies
homogeneity.) In this metric, as in the Schwartzschild metric,
there is, strictly speaking, a special observer to whom the Universe
is perfectly isotropic. We may or may not be this special oberver.
However, since the observed cosmic microwave background radiation and
$\Omega_{0}(r)$ look almost isotropic to us, we cannot be too far
from this oberver.
Since the left-hand side of the generalized Einstein equation has
$r$ dependence through the Ricci tensors,
the second ansatz is necessary in order to have the $r$ dependence
on the right-hand side to counter that of the left-hand side.

Having generalized both the Robertson--Walker  metric and the
Einstein equation, we now proceed to discuss their
consequences. When the non-vanishing elements of Ricci tensor
calculated
from Eq.(1) are substituted into the generalized Einstein equation
in Eq.(2),
we obtain the following equations for $tt$, $rr$, and $\theta \theta$
components :
\begin{eqnarray}
3 \frac{\dot{R}^2(t,r)}{R^2(t,r)}-2\frac{R''(t,r)}{R^3(t,r)}
+\frac{R'^2(t,r)}{R^4(t,r)}-4\frac{R'(t,r)}{rR^3(t,r)}
&=&8 \pi [G\rho](t,r) \\
2 \frac{\ddot{R}(t,r)}{R(t,r)}+\frac{\dot{R}^2(t,r)}{R^2(t,r)}
-\frac{R'^2(t,r)}{R^4(t,r)}-2\frac{R'(t,r)}{rR^3(t,r)}
&=& -8 \pi [Gp_r](t,r) \\
2 \frac{\ddot{R}(t,r)}{R(t,r)}+\frac{\dot{R}^2(t,r)}{R^2(t,r)}
-\frac{R''(t,r)}{R^3(t,r)}
+\frac{R'^2(t,r)}{R^4(t,r)}-\frac{R'(t,r)}{rR^3(t,r)}
&=&- 8 \pi [Gp_{\theta}](t,r),
\end{eqnarray}
where dots and primes denote, respectively, derivatives with
respect to $t$ and $r$, and  $p$ is the pressure. The equation
containing $ p_{\varphi}$ is identical to Eq.(5) with $p_{\theta}$
replaced by $p_{\varphi}$.
One more non-vanishing Ricci tensor $R_{01}$ is proportional to
momentum density
$T_{01}$ which we assume to be small
but finite. This nonvanishing $T_{01}$ has been, in the matter-dominated era,
responsible for allowing matter to flow from
its initially homogeneous distribution to the present
inhomogeneous distribution.
In addition, it also prevents $R(r,t)$ from being
factored out as $a(t)S(r)$, which is the case of no interest to us. As
designed, when the
$r$ dependence is simply $dropped$ (i.e., $R' = R'' =0$ ) and
$p_{r}=p_{\theta}=p_{\varphi}=p$ ,
we recover the Friedman equation with $k$=0.
With $p_{r}=p_{\theta}=p_{\varphi}=p \not= 0$,
we obtain, from Eqs.(4) and (5), a constraint on $R(t,r)$ given by
\begin{equation}
R''(t,r)-2\frac{R'(t,r)^2}{R(t,r)}-\frac{R'(t,r)}{r} =0~~.
\end{equation}
The constraint in Eq.(6) is valid
for $r \not= 0$,
because of its singular behavior at $r=0$.
In the following, $r\simeq 0$ implies very small  $r$, but
excluding
$r=0$, corresponding to our local neighborhood of 10 $\sim$ 50 Kpc.
The above constraint can easily be solved, yielding
\begin{equation}
R(t,r)=\frac{a(t)}{1-{\lbrack \frac{r a(t)}{2 A(t)} \rbrack}^2},
\end{equation}
where $a(t)$ and $A(t)$ are unknown functions of $t$ alone.
We have chosen a negative sign in Eq.(7) in order to have a locally
open Universe (see Eq.(8) below).
Different $t$ dependences in $A(t)$ and
$a(t)$ are protected by $p \not = 0$ and $T_{01} \not =0$.
The $R(t,r)$ in Eq.(7) has a singularity at $r=2A(t)/a(t)$. This
suggests that our Universe appears to be inside a cavity or bubble
surrounded by a wall with infinite density. The observed
approximate isotropy of the matter distribution and the cosmic microwave
background radiation indicates that we are not near the
wall. The wall may be much farther than our horizon. Such a model was
previously discussed [7] in an entirely
different context.

When Eq.(7) is substituted into the last three terms in Eq.(3),
Eq.(3) becomes
\begin{equation}
\left[  \frac{\dot{R}(t,r)}{R(t,r)} \right]^2
=\frac{8\pi G \rho}{3}+ \frac{1}
{A(t)^2}.
\end{equation}
(When $A(t)^2$ is constant, the last term in Eq.(8)
is nothing but a Cosmological Constant.
Thus the Cosmological Constant is generated,
even though we started with a generalized Einstein equation without it.)
Since we are interested in the matter-dominated era, we do not
consider equations with $p$ and $T_{01}$, assuming they are small,
but Eq.(8) is exact.
In any gravitational experiment or phenomenon,
a relevant quantity is always a product of $G$ and $\rho$ which
cannot be separated. We may define $G(r) \equiv G_N[1+\delta(r)]$
as an {\it effective \/} gravitational constant,
or interpret  $\rho$ as
$\rho(t,r)\simeq \overline{\rho}_{0} [R(t_0,r)/R(t,r)]^3[1+\delta(r)]$,
approximately valid in the matter-dominated era, with the present
local density $\overline{\rho}_{0}$.
Although both interpretations yield the same phenomenology,
their  physical implications are different. The former
is the $running$ of the gravitational constant, the latter
implying that the dark matter content increases as scale increases.
The truth may be some combination of the two.
Since our formulation is phenomenological, we cannot
distinguish between the two, but it is  extremely important
to separate them, for the difference reveals  the true
content of dark matter.

We now examine the physical meaning of Eq.(8). If we naively set $r=0$
in Eq.(8), we would end up with an equation which is not
only
different from the Friedman equation with $k =0$ but also simply
wrong.  Equation (8) in fact approaches the Friedman equation with
$k=0$ as
$r$ increases.  As cosmic scale grows at any fixed time,
the $ G\rho$ term also grows,
whereas the $1/A(t)^2$ term remains fixed, making
the curvature-like term
relatively  smaller compared with the $G\rho$ term.
This is the way we recover a flat Friedman Universe. (This is of course based
on the
assumption that the inflationary scenario is correct and we have
formulated accordingly.)

In this cosmology, we have
two expansion rates;
$H(t,r) \equiv {\dot{D}(t,r)}/{D(t,r)}$ and
${\cal{H}}(t,r) \equiv {\dot{R}(t,r)}/{R(t,r)}$.
The former is the $proper$ expansion rate with $D(t,r) \equiv \int _0 ^r
R(t,r') dr'$ and the latter is the one that appears in Eq.(8), which for
distinction we call the $scale$ expansion rate. We adapt the notation
$\overline{Q}$
for a quantity $Q(r)$ evaluated at our local neighborhood  $ r
\simeq 0$.
With this notation, we have
$ \overline{H}_0 =\overline{ \cal{H}}_0$.
Evaluated at $r\simeq 0$ and at present epoch, Eq.(8) becomes,
\begin{equation}
\overline {H}_0{^2} = \frac {8 \pi G_N \overline{\rho}_0}{3}
 +\frac{1}{A_0{^2}},
\end{equation}
which relates the unknown $A_0$ to $\overline{H}_0$ and $\overline
{\Omega}_0$.

The $\Omega _0(r)$ is, from Eq.(8),
$\Omega_{0} (r) \equiv  {(G\rho)_0}/{(G\rho)_{c,0}}= 1- 1/
{{\cal H}_0{^2}A_{0}{^2}}$,
where $(G\rho)_{c,0}$ is defined by ${\cal H}(t_0,r)^2 \equiv 8 \pi
(G\rho)_{c,0}/ 3$.
Its local version  (i.e., evaluated at $r\simeq 0$) is,
with Eq.(9),
\begin{equation}
\overline{\Omega}_0 = 1- \frac {1}{\overline{H}_0{^2}A_0{^2}}
     =\frac{1}{\overline{H}_0^2}\frac{8\pi G_N\overline{\rho_0}}{3}~.
\end{equation}
{}From Eqs.(8), (9), (10) and expression of $G\rho$ below Eq.(8), we obtain
\begin{equation}
{\cal H}_0(r) = \overline{H}_0 \sqrt{1+ \overline{\Omega}_0 \delta
(r)}~~.
\end{equation}
Similarly, we find
\begin{equation}
\Omega_{0}(r) = \overline{\Omega}_0 \frac{1 + \delta(r)}
{1 + \overline{\Omega}_{0} \delta(r)}.
\end{equation}

We note that this $\Omega_{0}$ is different from the standard
one defined with a constant $\rho_{c}$.
Both ${\cal {H}}_{0}(r)$ and $\Omega_{0}(r)$ are increasing functions of $r$
because $\delta (r)$ also is, as expected from the behavior of the
observed $\Omega_{0}(r)$. $\Omega_{0}(r)$  approaches unity, but
can never be greater than one. (The possibility of running $H_{0}$
was previously mentioned in [8].)
 The expressions given in Eqs.(11) and (12) are identical to  those in [1].

Now let us calculate the present age of the
Universe, $t_0$. In the age equation, the main $r$-dependent
contribution comes from
the $8 \pi G\rho$/3 term. Since $R(t,r)$ is a very slowly varying
function of $r$ up to scales of our interest, 15 Mpc,
defining $x \equiv R(t,r)/R(t_0,r)$,
and using Eqs.(8), (9), and (10),  and the approximate expression
for $\rho(t,r)$ mentioned below Eq.(8), we obtain the following age equation
\begin{equation}
t_0(r) \simeq \frac {1}{\overline{H}_0} \int_0 ^1 \frac{\sqrt{x}dx}
{\sqrt{\overline{\Omega}_0[1+ \delta(r)] +
x[1-\overline{\Omega}_0]}}.
\end{equation}
Equation (13) is identical to that of [1],
with exception of $\delta(r)$ being replaced by $\delta_G$ due to
the running $G(r)$ alone.
It is clear that $t_0(r)$ is now a decreasing function of $r$, with
the oldest age at $r\simeq 0$(or $\delta(r) \simeq 0$)
\begin{equation}
\overline{t}_0 \simeq \frac{0.9}{\overline{H}_0}= 18 ~~
\mbox{Gyr~~~  for
} ~~ \overline{H}_{0} \simeq 50 ~~\mbox{Km/secMpc}~~.
\end{equation}
This age can easily accommodate the observed ages, 14 $\sim$ 18 Gyr,
 of the oldest stars and globular clusters in our galaxy.

The proper expansion rate is given by, with Eqs.(7), (9), and (11),
\begin{equation}
H_0(r) = \overline{H}_0 \frac { \int _0 ^r {\sqrt{1+
\overline{\Omega}
_0 \delta(r')}}( 1- \beta r'^2)^{-1}dr'}
{\int _0 ^r (1-\beta r'^2)^{-1} dr'}~,
\end{equation}
where $\beta \equiv (a_0/A_0)^2/4$ is an unknown constant.
$H_0(r)$ is a weighted average of
${\cal H}_{0}(r)$ with weight $R(t_0,r)$, and is more slowly
increasing function of $r$ than that of Eq.(11) but approaches
 that of Eq.(11) for large $r$.

Before presenting numerical results, we mention that because of its specific
form, $R(t,r)$
dictates the form of the $r$ dependence of cosmological
quantities. This leads to
\begin{equation}
\delta(r)= \frac{1}{\overline{\Omega}_{0}}[(1 + \frac{\alpha r^2}{1-
\beta r^2})^2 -1]~~,
\end{equation}
where $\alpha$ and $\beta$ are positive parameters to be determined.
By fitting the (very uncertain) data of $\overline{\Omega}_{0}[1+\delta(r)]$ in
[6],
we obtained  $\alpha \sim 10^{-10}$ and
$\beta \leq 10^{-10}$ with $r$ converted into a proper
distance in units of Kpc. (Unfortunately, it is not possible to determine
$\beta$ because of
the poor data quality.)

Using Eq.(13) and the $\delta(r)$ as described above,
we obtain the age at 15 Mpc,
where one of the new measurements of $H_0$
has recently been made, as $t_0(15~Mpc) \sim  17$ Gyr.
Although this value agrees with the local ages,
it should $not$, in principle, be compared with
the local ages, as in the case of values
of the fine structure constant at  different energies.
We find, from Eq.(15),
$H_0(15{\rm{Mpc}}) \simeq  52$ Km/secMpc with
${\overline{H}}_0 = 50$ Km/secMpc.
Although, the calculated $H_0$ shows
an increasing trend, the increase is not fast enough to explain
the observed value,
$H_{0,\mbox{expt}}(15{\rm{Mpc}}) =(87 \pm 7)$ Km/secMpc [3].
The other measurements	 of $H_0$
based on the type Ia supernovae are
$H_0(B) = 52 \pm 8 $Km/secMpc and $H_0(V) = 55 \pm 8$ Km/secMpc [4].
These values were obtained from the Hubble diagrams in B and V
for the supernovae Ia
with distances ranging from $\sim$ 20 Mpc to $\sim$ 100 Mpc.
In this case, our calculated values of $H_0$ are 55 $\sim$ 70 Km/secMpc.
It remains to be seen whether this discrepancy will persist in the future
observation.  Comparison of the observed $\Omega_{0}$ and the one
predicted purely by quantum gravity can reveal the true content of dark matter.
Such a study and derivations of new (complex) relationships among distance,
redshift, and the two expansion rates in this new cosmology, which may
shed some light on the discrepancy in $H_{0}$,  are
given elsewhere. An important prediction of this cosmology is a
sharp rise of $\rho(r)$ beyond 100 Mpc.

\acknowledgments

The authors would like thank B.~Bosco, A.~Bottino, R.~Casalbuoni,
D.~Dominici, G. Grunberg, M.~Im, L.~Lusanna, T.N.~Pham and T.N.~Truong
for helpful discussions. Special thanks are due to A.~Chakrabarti
and G.~Feldman who pointed out the importance of nonvanishing $p$ and $T_{01}$.
In particular, without help of G.~Feldman,
this paper could not have been in its present form.
One of the authors (A.S.) would like to thank the late R.~Pierluigi
for very interesting discussions. This work was
supported in part by the National Science Foundation, U.S.A.

\end{document}